# Solution blending preparation of polycarbonate/graphene composite: boosting the mechanical and electrical properties

Emanuele Lago,[a,b] Peter. S. Toth,[a] Giammarino Pugliese,[c] Vittorio Pellegrini,[a] and Francesco Bonaccorso[a,*]

[a] Istituto Italiano di Tecnologia, Graphene Labs, Via Morego 30, 16163 Genoa, Italy.

[b] Dipartimento di Chimica e Chimica Industriale, Università di Genova, Via Dodecaneso 31, 16146, Genoa, Italy.

[c] Istituto Italiano di Tecnologia, Nanochemistry Department, Via Morego 30, 16163 Genova, Italy.

* Corresponding authors: francesco.bonaccorso@iit.it.

## Abstract

We report on the preparation of polycarbonate-based graphene (PC/G) composites, by using a simple and scalable solution blending method to disperse single- (SLG) and few-layer (FLG) graphene flakes, prepared by liquid phase exfoliation (LPE), in the polymer matrix. A solvent-exchange process is carried out to re-disperse the exfoliated SLG/FLG flakes in an environmentally friendly solvent, *i.e.* 1,3-dioxolane, which is also used to dissolve the polycarbonate pellets, thus facilitating the mixing of the polymer dispersion with the SLG/FLG flakes. The loading of SLG/FLG flakes improves the mechanical and thermal properties, as well as the electrical conductivity of the polymer, reaching a +26 % improvement of the elastic modulus at 1 wt% loading, and an electrical conductivity $10^{-3}$ S $m^{-1}$ at 10 wt% with a percolation threshold achieved at 0.55 vol%. The as-prepared PC/G composite with the aforementioned reinforced properties can be a promising material for 3D printing-based applications.

## Introduction

One of the main and most promising applications involving graphene –the two dimensional allotrope of carbon– is as filler in polymer nanocomposites.[1–6] In fact, the unique physical properties of graphene,[7–9] *i.e.* mechanical (Young's Modulus of 1 TPa and tensile strength of 130 GPa),[10] electrical (conductivity up to $10^8$ S $m^{-1}$)[11], and thermal (conductivity of ~$5\times10^3$ W $m^{-1} \cdot K^{-1}$)[12] properties, coupled with a large specific surface area (theoretically predicted 2630 $m^2$ $g^{-1}$),[13] make it appealing for the composite materials production.[4,14] In fact, graphene-based polymer composites have already shown enhanced mechanical,[15,16] thermal,[17] and electrical[18] properties with respect to pristine polymer matrices.

However, large quantities of graphene are needed for its exploitation in the composite field,[1,4] especially in view of industrial scalability. Although several techniques are available for the production of high-quality graphene monolayers on a substrate,[19] such as micromechanical cleavage,[9] chemical vapour deposition (CVD),[20] and growth on SiC substrate,[21] those approaches are clearly not suitable for composite applications. Nowadays, large-scale production of single- (SLG) and few- (FLG) layers graphene flakes by liquid phase exfoliation (LPE) of pristine graphite[22–24] is amongst the preferred production routes for the use of graphene as filler in nanocomposites.[19,24] In fact, the method allows to obtain SLG and FLG flakes dispersed in a solvent or in powder form and it is also offering the possibility of scaling up.[4,19,24]

In a LPE process, graphene flakes are produced by exfoliation of natural graphite,[4,22] or graphite oxide, [19] or graphite intercalated compounds (GICs),[19,25] in a solvent medium by an external driving force, such as ultra-sonication.[19,26] The choice of the solvent for the exfoliation process is crucial.[22,24,27] Suitable solvents have to minimize the interfacial tension between the liquid itself and the graphene flakes, *i.e.* solvents with surface tension of ~40 mN/m[4,19,22,26,28] such as N-methyl-2-pyrrolidone (NMP),[22,29] N,N-dimethylformamide (DMF),[22,30] and *ortho*-dichlorobenzene (DCB).[22,31] However, these solvents are toxic[32] and have high boiling point, *i.e.* more than 150 °C.[33] An alternative route to the solvent exfoliation[24] relies on the use of either surfactants[34,35] or polymers[36,37] which aid the exfoliation of graphite in water[26] or low boiling-point solvents, such as ethanol,[38] tetrahydrofuran (THF),[39] and chloroform,[39] stabilizing the exfoliated flakes against re-aggregation.[4,26,34–37] However, the residual of either surfactants or polymer increases the inter-flake contact resistance.[4] After the ultra-sonication process, an ultra-centrifugation step is required to remove thick flakes and un-exfoliated bulk graphite[29,40,41] from the as-prepared dispersion. The most common procedure is the sedimentation-based separation (SBS),[29,41–44] which separates flakes on the basis of their sedimentation rate, *i.e.* the tendency of a particle (or a flake) to settle out in the solvent, in response to centrifugal force acting on them.[45] The possibility to use a purified dispersion, *i.e.* either graphene flakes of well-defined morphology, such as lateral size and thickness, or without the presence of thicker and/or un-exfoliated graphitic flakes is beneficial for its use in various polymer matrices.[4,22,24] Summarizing, the growing demand of the polymer-based graphene composites needs an efficient and scalable production route, requirements which are satisfied by using the LPE approach. However, the as-produced dispersion by LPE has to be purified to obtain a high percentage of SLG. This requires an ultra-centrifugal step. Considering these facts and the up-scalability difficulties of ultra-centrifugation, dispersions containing pristine SLG flakes have limited availability on the commercial market, opening the avenues for novel developments in the scalable production techniques of graphene.

Another key requirement, apart from the graphene flakes morphology, to improve the properties of the final composite material with respect pristine polymer, relies on the optimal dispersion of the graphene flakes in the polymer matrices. The molecular interactions between the graphene flakes and polymer chains are due to weak van der Waals forces,[28] π–π stacking,[6,28] and hydrophobic-hydrophobic interactions.[28] These interactions hinder efficient connections between the pristine graphene flakes and the polymer chains, so graphene flakes usually do not form homogeneous composites.[1,2,14] In contrast, the epoxide, hydroxyl, carbonyl, and carboxyl groups present on the basal plane of graphene oxide (GO)[46,47] and also partially in reduced graphene oxide (RGO)[48,49] can interact with the polymer chains. Therefore, their use as a filler in the polymer-based composites is widely reported in literature.[1,3,4,14] Nevertheless, the presence of these groups acts as defects, in addition to the structural defects due to the oxidation process in the structure of the flakes.[48,50] The presence of such defects reduces the mechanical and electrical properties of GO and RGO flakes respect to graphene flakes,[48,50] *e.g.* GO is an insulator and has a Young's Modulus ranging between 200 and 600 GPa.[1,50,51] Besides, the dispersion of the flakes inside the polymer matrix is strongly dependent on the processing techniques used for the production of composite itself. Some of the most common processes for polymer composite preparation are melt blending,[52,53] solution blending,[54,55] and *in situ* polymerization.[56] Melt blending is industrially attractive due to its scalability and low-cost, while solution blending provides better mixing than melt blending between the exfoliated graphitic flakes and polymer matrix.[57–59]

Polycarbonate (PC) is a thermoplastic polymer with high mechanical stiffness (2.0–2.4 GPa)[60] and optical transparency (over 80% in visible spectrum, with a refractive index of 1.59),[60] and can be used in a wide range of applications: ranging from automotive and aeronautic industries,[60] to data storage (DVDs and CDs),[61] replacing glass,[61] and photonics.[62,63] Polycarbonate could also be used as 3D printer filament,[64] with higher mechanical and thermal properties with respect to polylactic acid (PLA)[65] and acrylonitrile-butadiene-styrene (ABS),[66] which are the reference materials used in these applications.[64] For instance, the heat deflection temperature, *i.e.* the temperature at which a polymer deforms under a specified load, for PC is in the 135–145 °C range,[61] whereas for PLA is ~60 °C,[65] while for ABS is ~100 °C.[67] This makes PC suitable for high-temperature required applications. Moreover, the exploitation of ABS or PLA filaments for 3D printing technology has also environmentally implications,[68,69] as ultrafine, <100 nm, particle emissions from melt during a 3D printing process is reported.[69] 3D printable graphene-based composites have been firstly reported in the case of PLA[70] and ABS.[70] In order to exploit the aforementioned properties of PC-based graphene composites for their use in 3D printing, further studies are needed. Currently, to the best of our knowledge, there is not reported any PC/graphene (PC/G) composite where LPE of natural graphite has been exploited, whereas primarily graphene derivatives such as GO or RGO have been previously used.[53,71–74] In particular, for what concerns the solution blending approach, the solvents used for the dissolution of PC are chloroform[72] and THF,[71] both suspected of being carcinogenic substances.[75,76]

In this work, we developed a simple solution blending process, to produce PC/G composite pellets using a 1,3-dioxolane-based dispersion, having a twofold function, acting as a dispersant for the graphene flakes and able to dissolve the PC for the realization of the final polymer composite. Structural characterization is carried out on both graphene flakes dispersion and PC/G composite samples using atomic force, transmission and scanning electron microscopies, and Raman spectroscopy. The LPE process produced prevalently SLG and FLG flakes with lateral size in range 200 – 600 nm, and thickness in the 0.7 – 1.4 nm range for 60-70% of the flakes, respectively. Mechanical tests and electrical conductivity measurements are carried out to investigate the effect of different loadings of SLG and FLG flakes on the composite's physical properties. If compared to the bare polymer, the as-produced PC/G composite shows a 26 % increment in Young's Modulus at 1 wt% loading and an electrical conductivity of ~$10^{-3}$ S m$^{-1}$ at 10 wt%, the latter representing a seven order of magnitude increment with respect to the pristine polymer.

## Experimental part

### Materials.

Polycarbonate pellets (MAKROLON®2405), natural graphite flakes (+100 mesh, 75% min), N-Methyl-2-pyrrolidone (NMP, Reagent Plus®, 99%) 1,3-dioxolane (Reagent Plus®, 99%) and acetone (ACS Reagent, ≥99.5%) are purchased from Sigma-Aldrich and used as received.

### Production of the graphene dispersion in 1,3-dioxolane

Graphene dispersion is produced by LPE of natural graphite. 500 mg of graphite flakes are dispersed in 50 mL of NMP and exfoliated in a sonic bath (VWR Ultrasonic Cleaner USC-THD) for 6 hours. Then, the dispersion is ultra-centrifuged at 10000 rpm (~17000 g), using SBS to remove un-exfoliated and thicker graphite flakes, for 30 min at 15 °C with an ultra-centrifuge (Beckman Coulter Optima™ XE-90, equipped with a SW32Ti rotor). Finally, the supernatant is collected by pipetting. A solvent exchange process is carried out for the re-dispersion of the exfoliated flakes in 1,3-dioxolane, a non-toxic and low boiling point (78 °C)[77] solvent, using a Heidolph Hei-Vap rotary evaporator. After the evaporation process of NMP, the graphitic material is collected and washed three times with acetone to remove the NMP residual using a compact centrifuge (Sigma-Aldrich). The washing step is repeated and the flakes are eventually dispersed in 50 ml of 1,3-dioxolane, adjusting its concentration at 10 mg mL$^{-1}$.

## Graphene dispersion characterization

Graphene dispersion is characterized morphologically (measuring lateral size and thickness) performing Transmission Electron Microscopy, TEM (Joel JEM 1101) and Atomic Force Microscopy, AFM (Bruker Innova®) on the samples once drop-casted to holey carbon coated copper grids and Si/SiO$_2$ wafers, respectively. Lateral size and thickness statistics are performed measuring ~100 flakes from both TEM and AFM images, respectively. Statistical analysis are fitted with log-normal distributions.[78,79] Raman spectra of SLG and FLG dispersion, drop casted on Si/SiO$_2$ wafers, are carried out using a Renishaw inVia confocal Raman microscope, with an excitation line of 532 nm (2.33 eV), a 50× objective and an incident power of 1 mW. For the statistical analysis 15 spectra have been collected and peaks are fitted with the Lorentzian function.

## Composite preparation

The PC/G composite is produced by exploiting the solution blending technique. Fig. 1 shows a schematic representation of the composite preparation following a three-step procedure. In the first step, PC pellets are dissolved in 1,3-dioxolane (15 wt/vol%) by stirring for 3 h using a magnetic stirrer and then mixed with SLG and FLG dispersion in the same solvent (Fig. 1a). As second step, SLG and FLG dispersion in 1,3-dioxolane (10mg/mL) is added at different weight fractions, namely between 0.0 wt% and 3.0 wt% with respect to PC weight. The resulting dispersions are then mixed by means of ultra-sonication for 2 h, maintaining the temperature in the range 25–40 °C, in order to obtain a thorough mixing. In the third-step the composite dispersion is coagulated/precipitated forming pellets by pouring water (Fig. 1b), and then dried in a vacuum oven (Binder VDL 115) at 80 °C for 12 h.

## Transmission electron microscopy (TEM)

Graphene, $h$-BN, MoS$_2$ and WS$_2$ are prepared by drop casting dispersions onto ultrathin C-film on holey carbon 400 mesh Cu grids, from Ted Pella Inc. The graphene samples are diluted 1:50, while the $h$-BN, MoS$_2$ and WS$_2$ samples are diluted 1:20. The grids are stored under vacuum at room temperature to remove the solvent residues. TEM images are taken by a JEOL JEM-1011 transmission electron microscope, operated at an acceleration voltage of 100 kV. High-resolution TEM (HRTEM) is performed using a 200 kV field emission gun, a CEOS spherical aberration corrector for the objective lens, enabling a spatial resolution of 0.9 Å, and an in-column image filter (Ω-type). Finally, the pellets, previously heated up to 225 °C and pressed at 2 t for 5 min, are compression moulded to form films, using a Madatec Atlas T8 press.

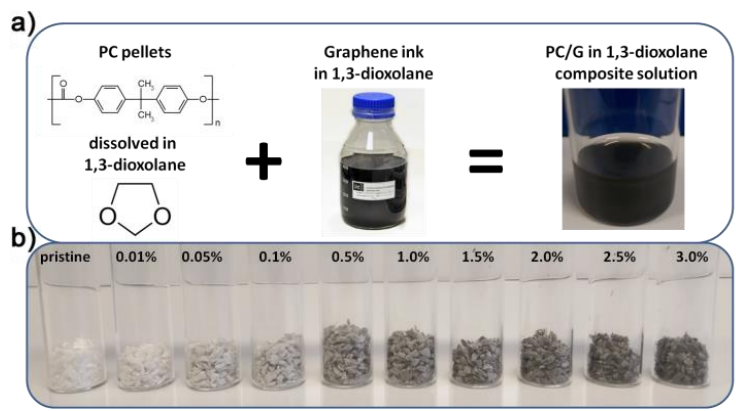

Figure 1. (a) Scheme of solution blending of PC/G composites. (b) Photograph of PC/G composite pellets at different loading of graphene-based flakes.

### Characterization of the composite material

The composite morphology is investigated by using Scanning Electron Microscopy (SEM) with a FE-SEM Joel GSM-7500FA. Raman spectra of composite material samples are acquired using a Renishaw inVia confocal Raman microscope, using an excitation line of 532 nm (2.33 eV) with a 50× objective and an incident power of 1 mW. Peaks are fitted with Lorentzian functions. Thermogravimetric analysis (TGA) is carried out with a TGA Q500-TA Instrument. During TGA, samples are heated from 30 °C to 800 °C at a heating rate of 10 °C min$^{-1}$ under nitrogen atmosphere set at a flow rate of 50 mL min$^{-1}$. Mechanical properties of composite films are measured using an Instron dual column table top universal testing System 3365, with 5 mm min$^{-1}$ cross-head speed. The tensile measurements are carried out on five different specimens for each film according to ASTM D 882 Standard Test Methods for Tensile Properties of Thin Plastic Sheeting. The electrical conductivity (EC) is measured on 1 cm × 0.5 cm pieces of the compression moulded films. The sheet resistance and as-calculated DC conductance values of the pristine polymer and composite materials is recorded using a four-point probe and source meter (Keithley, 2612A) in DC regime (samples are placed crossing on four parallel palladium wires spaced by 1 mm).

## Results and discussion

### Characterization of graphene dispersion in 1,3-dioxolane

In order to achieve a homogeneous dispersion of exfoliated graphene flakes into PC matrix, we exploited solution-blending technique for the production of PC/G composite.[57,58] In a common solution-blending process, both matrix and filler have to be dissolved in the same solvent and then mixed. For the production of graphene-based composites exploiting this technique, PC is usually dissolved in THF[72] or chloroform,[71] i.e., solvents having as a downside the toxicity.[75,76] Therefore, we selected 1,3-dioxolane as solvent for PC, because its low-toxicity, *i.e.* it is not a carcinogenic reagent if compared with NMP (see safety data sheet, SDS).[80] To the best of our knowledge this is the first time that both PC and graphene flakes are dispersed in 1,3-dioxolane for their solution blending. This solvent is not suitable for the direct LPE of graphite because its surface tension, 32.6 mN m$^{-1}$,[81] does not match the surface energy of graphene, i.e., 46.7 mN m$^{-1}$.[22,29] Thus, LPE of graphite is carried out in NMP,[22] by means of low-power ultra-sonication. NMP is one of the best solvent used for the dispersion of graphitic flakes because its surface tension is 41 mN m$^{-1}$,[82] close to the surface energy of graphene.[22,29] The result of the exfoliation process is a heterogeneous dispersion of thin/thick and small/large graphitic flakes.[19] This dispersion is then purified exploiting the SBS process.[41,44]

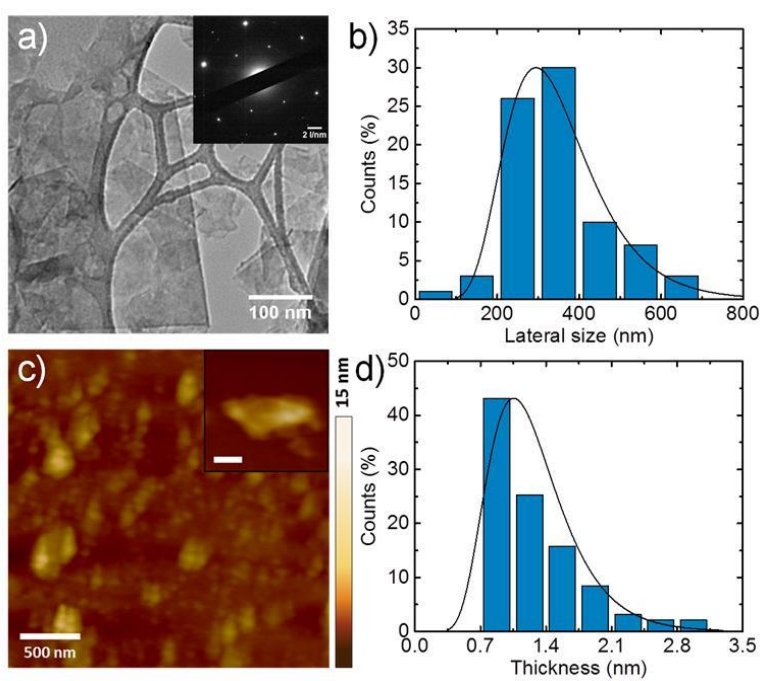

**Figure 2.** Morphological characterization of graphene dispersion in 1,3-dioxolane. (a) TEM image of graphene dispersion in 1,3-dioxolane; the electron DP image of a graphene flake is depicted in the upper inset. (b) Statistical analysis from TEM images of the flakes lateral size. (c) AFM image of graphene flakes casted from dispersion in 1,3-dioxolane; the upper inset shows a higher magnification image of a graphene flake (scale bar corresponds to 100 nm). (d) Statistical analysis **of the flakes thickness.**

In order to re-disperse the graphitic flakes, previously exfoliated in NMP, into 1,3-dioxolane, we performed a solvent-exchange process *via* rotary evaporation, see Experimental section. Graphene dispersion in 1,3-dioxolane is morphologically characterized by TEM and AFM (Fig. 2) and Raman spectroscopy (Fig. S1). TEM image of graphene flakes in 1,3-dioxolane is depicted in Fig. 2a. The electron diffraction pattern (DP) image (inset to Fig. 2a) confirms the typical lattice reflections present from a graphene flake, the expected honeycomb lattice, and crystallinity.[83] The statistical analysis of lateral sizes (Fig. 2b) shows a distribution of lateral size in the 200–600 nm range. The statistic is fitted by a log-normal distribution with a most probable value of lateral size peaked at 295 nm and standard deviation of 0.34. A representative AFM image and statistical analysis of graphene dispersion casted on a Si/SiO$_2$ wafer (see Experimental) are shown in Fig. 2c and Fig. 2d, respectively. More than 70 % of the measured thicknesses are in the range 0.7–1.4 nm, corresponding to 1–3 layers of graphene flakes, considering that the first layer on SiO$_2$ substrate has a height of ~0.7 nm.[84,85] The most probable value occurs at 1.05 nm (with a standard deviation of 0.35). The presence of defect-free SLG/FLG flakes is also confirmed by the Raman spectroscopy (see Appendix Fig. S1).Polyamide–12-graphene composite

The composite is prepared by melt blending. The as-produced WJM0.1 sample is dried using a rotary evaporator (Heidolph, Hei-Vap Value, at 70 °C, 5 mbar). Polyamide–12 (Sigma Aldrich) and the graphene WJM0.10 powder (1% in weight) are mixed in a twin-screw extruder (model: 2C12-45L/D, Bandiera) at 175°C. The mechanical properties of bare Polyamide–12 and Polyamide–12/graphene composites are measured using a universal testing equipment (Instron Dual Column Tabletop System 3365), with 5 mm/min cross-head speed. The tensile strength measurements are carried out on 7 different samples for each composite material according to ASTM D 882 Standard test methods.

## Morphological characterization of composite

Blend compatibility (*i.e.* expecting macroscopically uniform physical properties), filler dispersion, and the interfacial bond between the polymer matrix and the SLG-FLG flakes are investigated by means of SEM measurements on both the pristine PC and PC/G composite.

Scanning electron microscopy images of the cross section of pristine PC are shown in Fig. 3, at low (Fig. 3a) and high (Fig. 3b) magnifications, and PC/G composite of 3 wt%, at low (Fig. 3c) and high (Fig. 3d) magnifications, respectively. Graphene flakes (the brighter, angular shaped objects, marked by red arrows) can be clearly seen as uniformly dispersed in the polymer matrix, which is also confirmed by Raman characterization of the PC/G composite. Fig. 4a shows representative Raman spectra of graphene dispersion (trace black), polycarbonate (trace red) and PC/G composite dispersion after blending (trace blue). The Raman spectrum of the PC/G 3 wt% of graphene loading reveals a thorough mixing of the graphene flakes within the polymer matrix; indeed, there is the presence of peaks related to both SLG/FLG and the PC. The dotted lines referred as i), ii) and iii) correspond to C–CH$_3$ (~889cm$^{-1}$), C–O (~1235cm$^{-1}$) and ring stretches (~1606cm$^{-1}$) of PC,[86] respectively. Statistical analysis on Pos(2D) (Fig. 4b) shows a slight blue shift of the composite with respect to the one in pristine PC. In fact, Pos(2D) is in the range 2688–2700 cm$^{-1}$ for the dispersion and 2692–2702 cm$^{-1}$ for the PC/G composite. The blue shift is attributed to π–π interactions between the graphene flakes and the polymer.[1,87] However, the 2D peak in the composite still shows a Lorentzian line-shape differently from graphite (see FWHM(2D) distribution, Fig. 4c), indicating that FLG flakes are electronically decoupled.[41,44] The lack of correlation between $I_D/I_G$ and FWHM(G) (Fig. 4d) in the PC/G composite, as for the case of the graphene dispersion (see also Fig. S1f), suggests that no defects in the SLG and FLG flakes are introduced by the solution blending process.[88] More details of the Raman spectroscopy results and statistical analyses obtained investigating graphene dispersion in 1,3-dioxolane are found in the Appendix

Round-shape Cu disks (diameter of 1.5 cm, thickness of 25 µm, Sigma-Aldrich) are cleaned with acetone and 2-propanol (Sigma-Aldrich) in ultrasonic bath for 10 min. Then, the Cu disks are dried and weighted (Mettler Toledo XSE104). Subsequently, 100 µL of as-prepared WJM0.10 are drop-cast on each Cu disk under air atmosphere at 80 ˚C and then dried at 120 ˚C and 10$^{-3}$ bar for 12 hours in a glass oven (BÜCHI, B-585). The graphene mass loading (~1 mg) for each electrode is calculated by subtracting the weight of bare Cu foil from the total weight of the electrode.

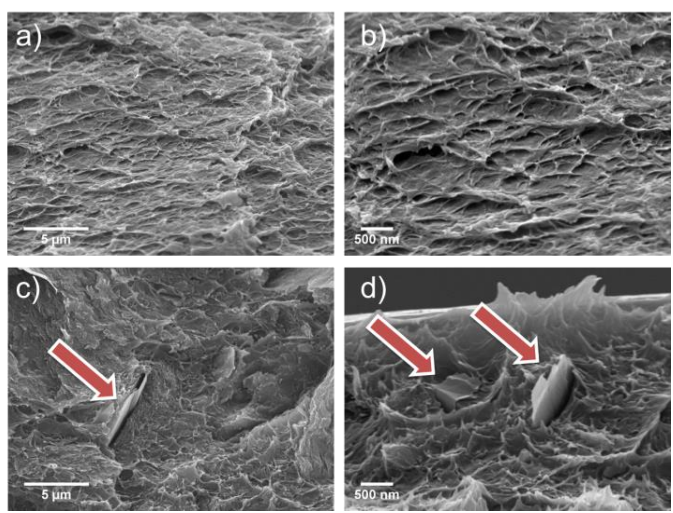

**Figure 3**. SEM images of cross section area of pristine PC, at (a) low and (b) high magnification, and PC/G composite of 3 wt% graphene loading, at (c) low and (d) high magnification. The red arrows mark the graphene flakes.

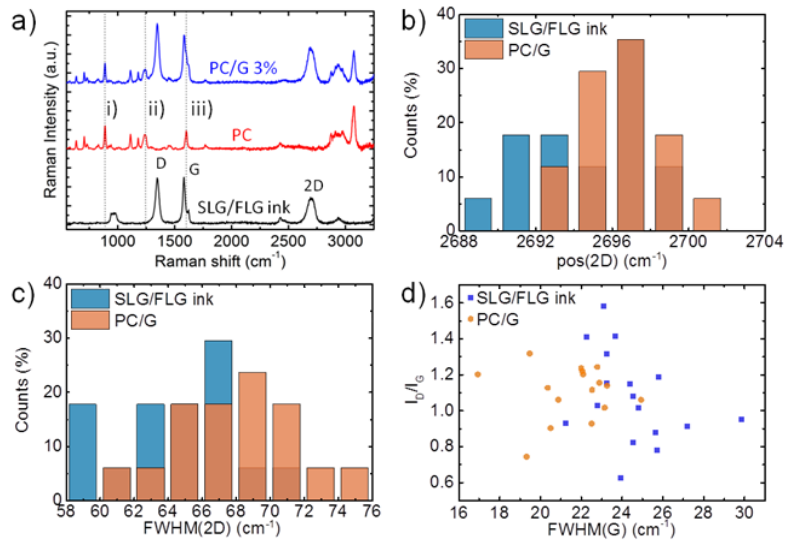

**Figure 4.** (a) Raman spectra of graphene dispersion (trace black), pristine PC (trace red) and polycarbonate/graphene dispersion at 3% of loading (trace blue), (b-d) statistical analysis on Raman spectra of graphene dispersion and composite samples: (b) Pos(2D), (c) FWHM(2D), (d) $I_D/I_G$ respect to FWHM(G).

## Composite enhanced performances

The thermal behaviour of a composite material in view of applications, such as for example 3D printing, has to be known.[70] Therefore, considering the application of the PC/G composite material for 3D printing using Fused Deposition Modelling (FDM) technology, these thermal behaviours are investigated. The FDM is an additive manufacturing technology commonly used for modelling, prototyping, and production of polymer-based objects, and it is one of the most common used techniques for 3D printing.[89,90] Using the FDM process, the thermoplastic polymer filament is heated above its glass transition temperature ($T_g$) and extruded through the nozzle of the 3D printer, then the printed material is cooled down to room temperature forming the product. The TGA and differential TGA (DTGA) analyses on PC (black curve) and PC/G at 1 wt% (red curve) are depicted in Fig. 5a and Fig. 5b, respectively. The first loss of weight in the case of both pristine PC and composite material occurs in the 100–150 °C range, corresponding to ~5 wt% loss. This decrease is attributed to the evaporation of residual solvents and/or small organic groups.[91] The pyrolysis, corresponding to the main loss of weight, starts for pristine PC at ~370 °C and is due to the cleavage of the carbonate groups,[91] whereas it is reduced in the composite because fillers reduce mobility of polymer chains, so the degradation temperature of the composite loaded with 1 wt% of SLG-FLG flakes, evaluated as the peak in the DTGA curve, increases of ~86 °C. The peak of DTGA, corresponding indeed to the maximum reaction speed in which the sample is degraded, is at ~421 °C for the pristine polymer and at ~507 °C for the PC/G composite 1 wt% loaded, showing that the presence of graphene flakes increases the thermal stability of the polymer. At 800 °C, residual chars of 20wt% of PC and 21wt% of composite are found, in agreement with the loading of filler.

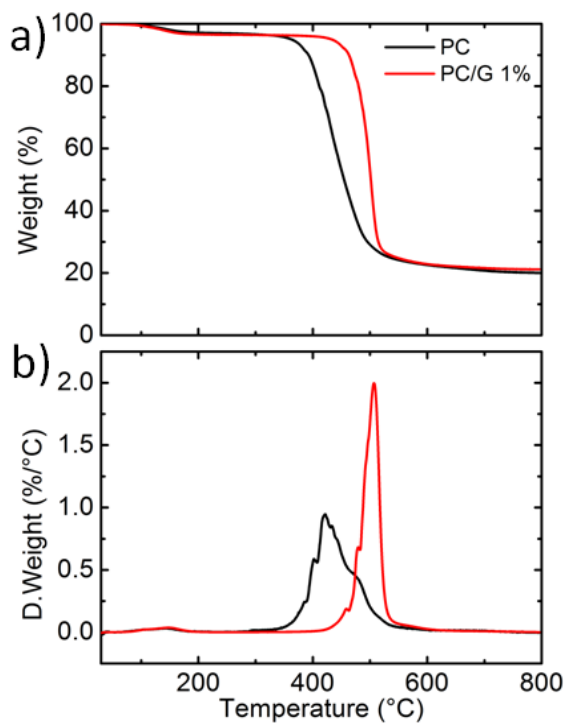

**Figure 5.** (a) TGA analysis of pristine PC (black curve) and PC/G 1 wt% (red curve). (b) DTGA analysis of pristine PC (black curve) and PC/G 1 wt% (red curve).

Stress *vs.* strain curves of pristine PC (black curve) and PC/G composite at 1 wt% (red curve) graphene loadings are shown in Fig. 6a. The Young's Modulus (E) (Fig. 6b), defined as the slope of the stress *vs.* strain curve in the elastic region, reaches a maximum value for the 1 wt% content of SLG/FLG flakes of 1455±28 MPa. The increment with respect to pristine PC, having a measured E value of 1151±44 MPa, is ~26%. Increasing the content of filler above 1 wt%, the E value decreases, reaching a minimum value of ~1353 MPa at 1.5 wt%, which, however, still corresponds to ~17% improvement with respect to pristine PC (1151±44 MPa).

The decreases of E with a loading of SLG and FLG flakes superior to 1 wt%, could be associated to the occurrence of agglomeration of the flakes,[55,92,93] although further studies to ascertain this phenomenon are needed. The presence of SLG and FLG flakes also improve the tensile strength at yield ($\sigma_y$, defined as the stress at which a material begins to deform plastically, Fig. 6c) and ultimate tensile strength ($\sigma_u$, defined as the higher stress value reached in the stress *vs.* strain curve, Fig. 6d) of the polymer. Contrariwise to the results obtained for E, where exceeding the 1 wt% of loading there is a sudden decrease, $\sigma_y$ and $\sigma_u$ remain almost constant, as a 'saturation/like' behaviour. This opposite behaviour is rooted in the linear elastic behaviour of plastics, as the E corresponds to the stiffness, rigidity of a sample, while the $\sigma_Y$, and $\sigma_U$ is the capacity of the material (to withstand loads tending to elongate), and the stress at which plastic deformation begins, respectively. The full mechanical characterization data are summarized in Table S1 (see Appendix).†

The increments in the mechanical properties that we obtained are higher with respect to other works involving the use of PC. For example, Kim *et al.*[53] reported a ~6.7% and ~20.7% increments of E at 1 wt% and 2.5 wt% of functionalized graphene sheets (thermally exfoliated graphite oxide) loading, respectively, while Shen *et al.*,[94] reported a 6.8% increment of E at 10 wt% of RGO loading. The same group[95] also reported a 72.1% increment of E using epoxy-functionalized GO, in which GO flakes were dispersed in DMF and PC dissolved in THF. However, both solvents have toxicity issues, as discussed above. Mittal *et al.*[96] reported enhancement of ~23% in E but with 7 wt% of RGO flakes loading.

We anticipate that the mechanical properties of our composite material could be additionally enhanced by optimizing the aspect ratio of the graphene flakes (lateral size *vs.* thickness), as reported in the case of PVA,[93] where the reported aspect ratio of flakes is ~1900 and the enhancement of E with respect pristine PVA is ~66% at 0.36 vol% (~0.65 wt%) of loading (as comparison, the aspect ratio of the graphene flakes used in this work is ~280). Considering the work of Coleman and co-workers,[93] it seems that the resulting mechanical enhancement of the polymer-based graphene composites, is rooted on the differences in the structural and morphological properties of the graphene flakes. However, further systematic studies are needed to verify this hypothesis.

Finally, static electrical conductivity measurements (DC regime) on composite as function of SLG/FLG content in the composite material (PC/G), are presented in Fig. 7. The PC/G composites with SLG and FLG concentrations up to 10 wt% are prepared for this characterization.

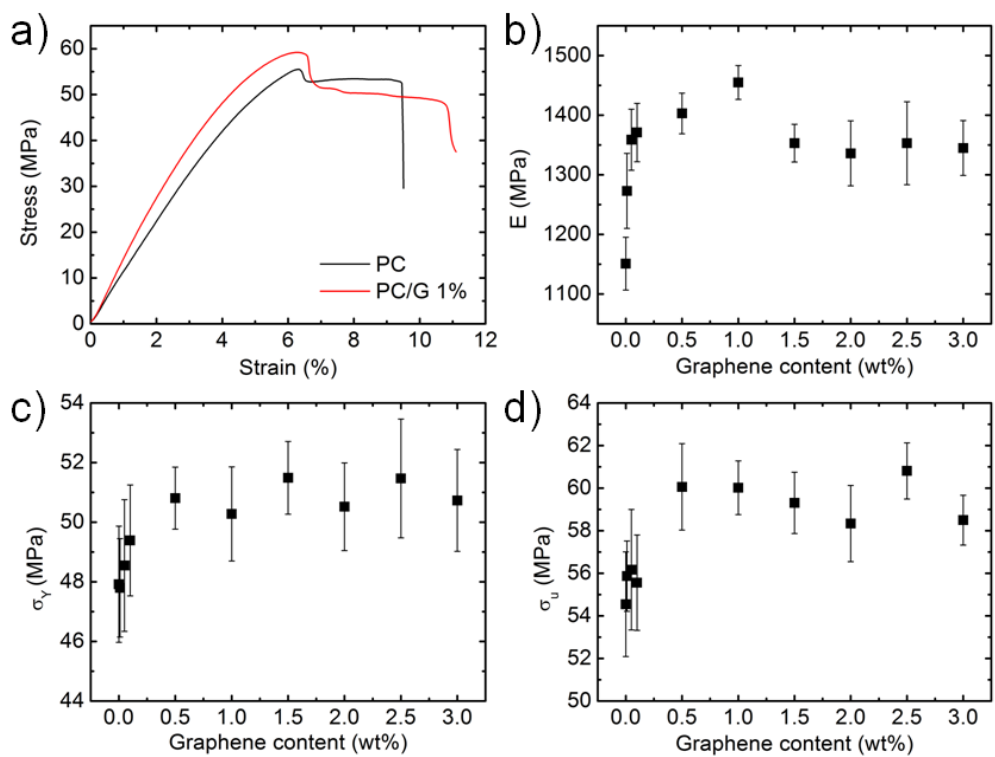

**Figure 6.** Mechanical properties of composites and pristine materials: (a) stress *vs.* strain curves, (b) Young's modulus (E), (c) tensile strength at yield ($\sigma_Y$), and (d) ultimate tensile strength ($\sigma_u$).

The pristine polymer matrix exhibits DC conductivity of the order of $10^{-11}$ S m$^{-1}$, reaching ~$10^{-7}$ S m$^{-1}$ for 2 wt% and almost $10^{-3}$ S m$^{-1}$ at 10 wt% of loading. The improvement on DC electrical conductivity of the composite follows a percolation behaviour,[18,71,97] with a percolation threshold ($\phi_c$) of ~1 wt% (~0.55 vol%). This value is in line with what achieved by other works involving PC/G composites.[53,71] The percolation threshold corresponds to the critical volume of nanoparticles, in this case SLG and FLG flakes, such that they are able to generate a conductive path for charge carriers.[98] When the content of filler exceeds $\phi_c$, there is a suddenly increase in DC electrical conductivity of the PC/G composite. According to the percolation theory, electrical conductivity of composite is related to volume fraction $\phi$ of filler by *Eq. 1*:

$$\sigma_{DC} = \sigma_0 \left[(\phi - \phi_c)/(1 - \phi_c)\right]^t \qquad (1)$$

for $\phi > \phi_c$, where $\sigma_{DC}$ is the DC electrical conductivity, $\sigma_0$ is referred as the conductivity of fillers, $\phi_c$ is the percolation volume fraction, and t is the critical exponent, which depends on the percolation model.[98,99] From the linear fit of $\lg(\sigma_{DC})$ vs. $\lg[(\phi-\phi_c)/(1-\phi_c)]$, shown as inset to Fig. 7, we found that, $\sigma_0 = 10^{(2.152\pm0.823)}$ S m$^{-1}$ = 141.91 ± 6.65 S m$^{-1}$ and t = 4.027 ± 0.411. The percolation threshold could be further lowered increasing the aspect ratio of the filler, facilitating the formation of a conductive path at a lower volume loading.[18,100,101]

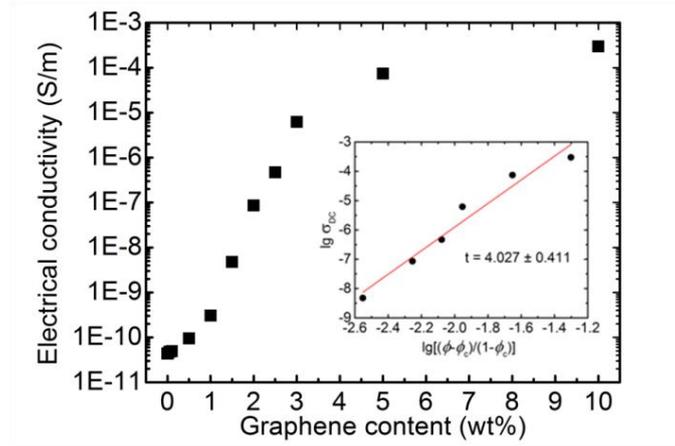

**Figure 7.** Electrical conductivity ($\sigma_{DC}$) of composite vs. SLG/FLG flakes content showing percolation behaviour. The plot of $\log(\sigma_{DC})$ vs. $\lg[(\phi-\phi_c)/(1-\phi_c)]$, where $\phi$ is the volume content and $\phi_c$ the percolation threshold, with a linear fit of experimentally measured points is shown in the inset.

Conclusions

In this study we proposed a scalable method for the production of a PC/G composite. Single and few-layers graphene flakes, used as filler, are obtained by means of LPE of pristine graphite.

  Exfoliation is carried out in NMP and then, by exploiting a solvent exchange process, the graphene flakes are dispersed in 1,3-dioxolane, which is a non-toxic, environmentally-friendly and low-boiling point solvent. The final composite is then obtained by mixing the graphene-based dispersion and the polymer solution in 1,3-dioxolane, by means of solution blending. Raman spectroscopy and SEM microscopy have shown the homogeneous dispersion of the single- and few-layer graphene flakes in the PC matrix.

The thermal stability of the PC/G containing 1 wt% of SLG and FLG flakes filler compared to the pristine PC is assessed by TGA analysis demonstrating an upshift of the degradation temperature of ~86 °C. Furthermore, a 26% improvement of Young's Modulus with respect to pristine PC is reached for 1 wt% content of SLG and FLG flakes. The presence of the fillers also enhances the yield- and ultimate- tensile strength. The PC/G composite electrical percolation threshold is found at ~1 wt% of graphene loading, reaching electrical conductivity of ~$10^{-3}$ S m$^{-1}$ at 10 wt%.

The as-produced PC/G composite can be the ideal starting material for 3D printing applications in constructing 3D electronics, due to its increased mechanical and electrical properties and thermal stability coupled with the environmentally-friendly approach here proposed.

We have demonstrated the wet-jet milling as a method to produce large quantities of few-layer graphene dispersions, achieving concentration up to 10 gL$^{-1}$ with an exfoliation yield, *i.e.,* ratio between the weight of the processed material and the weight of the starting graphite flakes, of 100%. Our lab-scale set-up enables a production capability of up to 2.35 L h$^{-1}$. The average time required to produce one gram of exfoliated graphite is 2.55 min (23.5 g h$^{-1}$), which favourably outperforms other liquid-phase exfoliation processes such as ultrasonication, high-shear exfoliation, or microfluidization. The exfoliated flakes have a lateral size of ~460 nm and a thickness lower than 4 nm. Further purification, by ultracentrifugation of the as-produced WJM0.10 sample, promotes the enrichment of single-layer graphene. In fact, the percentage of single-layer graphene passes from ~10% in the as-prepared WJM0.10 sample to ~57% in the purified one. Additionally, we have shown the feasibility of wet-jet milling for the exfoliation of inorganic layered crystals, *i.e.*, hexagonal boron nitride, molybdenum disulphide, and tungsten disulphide, obtaining flakes with lateral sizes of 380, 500, and 340 nm, respectively.

The as-produced graphene flakes can be used without further purification for added-value applications. In particular, we have demonstrated the as-produced WJM0.10 as active material for anodes in lithium ion batteries, reaching 420 mAh g$^{-1}$; as filler in Polyamide–12 composites, getting an improvement of 34% of the flexural modulus; as ink-jet printable conductive ink, obtaining state-of-the-art electrical conductivity of ~1.3 S cm$^{-1}$.

## Acknowledgements

The authors acknowledge the funding support from the European Union Seventh Framework Programme under grant agreement no. 604391 Graphene Flagship. E.L. also thanks the Educational Fund (Italia) for a PhD scholarship.

# Appendix

## Raman characterization of SLG/FLG flakes dispersion in 1.3-dioxolane

Figure S1 shows the Raman spectroscopy results obtained investigating graphene dispersion in 1,3-dioxolane, where a representative Raman spectrum is depicted in Fig. S1a. The Raman fingerprints of graphene are the G (~1580cm$^{-1}$) and 2D (~2700cm$^{-1}$) peaks.[1–4] If graphene flakes have defects, a D peak (~1350cm$^{-1}$) also appears.[1–4] The G peak corresponds to the $E_{2g}$ phonon at the Brillouin zone center.[2] The D peak is due to the breathing modes of *sp$^2$* rings and requires defects for its activation, the 2D peak is the second order of the D peak and is always visible, even without the presence of defects.[5,6] Statistical analysis shows 2D peak position (Pos(2D)) (Fig. S1b) in the 2688 – 2700 cm$^{-1}$ range. The analysis of 2D peak in position, width (full width at half maximum, FWHM, Fig. S1c) and intensity (respect to the G peak, $I_{2D}/I_G$, Fig. S1d), gives information about the number of layers of the graphene flakes.[1–3] The FWHM (2D) is in average ~70 cm$^{-1}$, while the $I_{2D}/I_G$ ratio is higher than 0.5, which represents the reference value of graphite.[1] These results suggest that the dispersion is composed by a combination of both single-(SLG) and few-layer graphene (FLG) flakes,[3,7] in agreement with the atomic force microscopy (AFM) data reported (see Fig. 2d in main text). Raman spectroscopy allows also to provide indication about the nature of defects in the graphene flakes.[1,2,7] In fact, by combining the $I_D/I_G$ ratio with FWHM(G) allows us to discriminate between disorder localized at the edges and disorder in the bulk. In the latter case, a higher I(D)/I(G) would correspond to higher FWHM(G). The I(D)/I(G) ratio is in the range 0.6 – 1.6 (Fig S1e), but the lack of correlation between I(D)/I(G) and FWHM(G) (Fig. S1f) proves that the major contribution to the D peak comes from the sample edges (see Fig. 2b in main text) rather than to the presence of structural defects.[5,7] Moreover, in the high-defect concentration regime FWHM(G) and FWHM(D') become broader and eventually merge into a single band.[5,7]

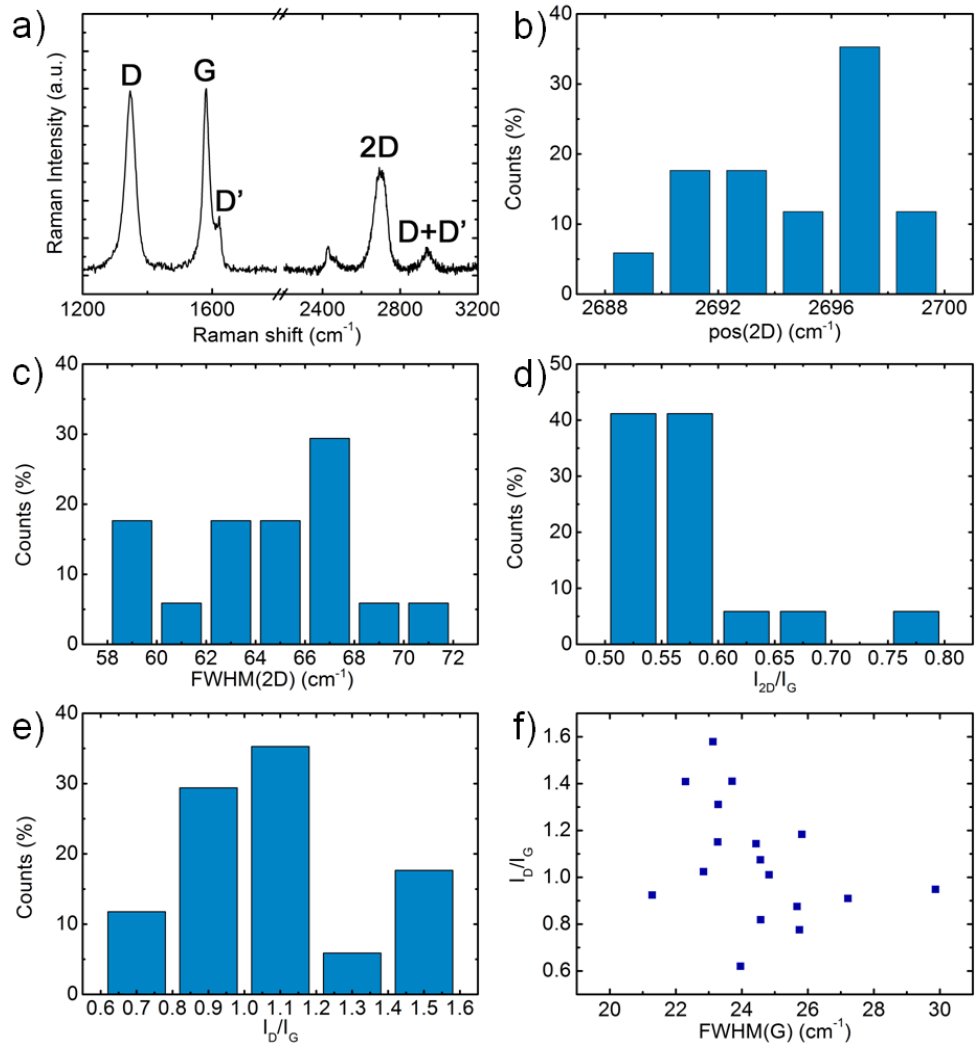

**Figure S1.** (a) Representative Raman spectrum of graphene flakes dispersed in 1-3 dioxolane. (b – f) Statistic analysis on the acquired Raman spectra: (b) pos(2D), (c) FWHM(2D), (d) $I_{2D}/I_G$, (e) $I_D/I_G$, and (f) $I_D/I_G$ as a function of FWHM(G).

### The mechanical characterization data of polycarbonate/graphene composites

The full mechanical characterization data are summarized in Table S1. The increments in mechanical properties are stated as $\Delta E$, $\Delta\sigma_Y$, and $\Delta\sigma_u$.

Table S1. Summary of mechanical properties of polycarbonate/graphene composite.

| SLG/FLG content | E | error E | ΔE | $\sigma_Y$ | error $\sigma_Y$ | $\Delta\sigma_Y$ | $\sigma_u$ | error $\sigma_u$ | $\Delta\sigma_u$ |
|---|---|---|---|---|---|---|---|---|---|
| wt% | MPa | MPa | % | MPa | MPa | % | MPa | MPa | % |
| 0.00 | 1151 | 44 | — | 47.9 | 2.9 | — | 55.2 | 3.4 | — |
| 0.01 | 1273 | 62 | 10.6 | 47.8 | 2.1 | -0.2 | 55.9 | 2.5 | 1.3 |
| 0.05 | 1359 | 51 | 18.1 | 48.5 | 2.4 | 1.3 | 56.2 | 3.8 | 1.8 |
| 0.10 | 1371 | 48 | 19.1 | 49.4 | 3.5 | 3.0 | 55.6 | 6.0 | 0.7 |
| 0.50 | 1403 | 34 | 21.9 | 50.8 | 1.0 | 6.0 | 60.1 | 0.9 | 8.9 |
| 1.00 | 1455 | 28 | 26.4 | 50.3 | 1.5 | 4.9 | 60.0 | 1.1 | 8.8 |
| 1.50 | 1353 | 31 | 17.6 | 51.5 | 0.2 | 7.4 | 59.3 | 1.4 | 7.6 |
| 2.00 | 1226 | 54 | 6.5 | 50.5 | 0.4 | 5.4 | 57.3 | 2.4 | 4.0 |
| 2.50 | 1353 | 69 | 17.4 | 51.6 | 1.9 | 7.4 | 60.8 | 0.2 | 10.3 |
| 3.00 | 1345 | 46 | 16.8 | 50.7 | 1.7 | 5.8 | 58.5 | 1.1 | 6.1 |